\documentclass[ba]{imsart}

\usepackage{booktabs}
\usepackage{comment}
\usepackage[noend]{algpseudocode}
\usepackage[ruled,vlined]{algorithm2e}
\usepackage{tablefootnote}

\usepackage{graphicx}

\RequirePackage{amsthm,amsmath,amsfonts,amssymb}
\RequirePackage[authoryear]{natbib}
\RequirePackage[colorlinks,citecolor=blue,urlcolor=blue,backref=page,backref=page]{hyperref}
\hypersetup{colorlinks,citecolor=blue,urlcolor=blue,filecolor=blue,backref=page}
\RequirePackage{graphicx}

\pubyear{2025}
\arxiv{TBA}
\volume{TBA}
\issue{TBA}
\firstpage{1}
\lastpage{1}

\startlocaldefs
\theoremstyle{plain}

\theoremstyle{definition}

\theoremstyle{remark}

\endlocaldefs

\begin{document}

\begin{frontmatter}
\title{BARTSIMP: flexible spatial covariate modeling and prediction using Bayesian additive regression trees}
\runtitle{BART for Spatial INLA Modeling and Prediction}

\begin{aug}
\author[A]{\fnms{Alex Ziyu}~\snm{Jiang}\ead[label=e1]{jiang14@uw.edu}\orcid{0009-0002-7733-2612}}
\and
\author[B]{\fnms{Jon}~\snm{Wakefield}\ead[label=e2]{jonno@uw.edu}}
\address[A]{Department of Statistics, University of Washington, US\printead[presep={,\ }]{e1}}

\address[B]{Department of Statistics and Biostatistics, University of Washington, US\printead[presep={,\ }]{e2}}
\runauthor{A. Jiang and J. Wakefield}
\end{aug}

\begin{abstract}
Prediction is a classic challenge in spatial statistics and the inclusion of spatial covariates can greatly improve predictive performance when incorporated into a model with latent spatial effects. It is desirable to develop flexible regression models that allow for nonlinearities and interactions in the covariate specification. Existing machine learning approaches that allow for spatial dependence in the residuals fail to provide reliable uncertainty estimates. In this paper, we investigate the combination of a Gaussian process spatial model with a Bayesian Additive Regression Tree (BART) model. The computational burden of the approach is reduced by combining Markov chain Monte Carlo (MCMC) with the Integrated Nested Laplace Approximation (INLA) technique. We study the performance of the method first via simulation. We then use the model to predict anthropometric responses in Kenya, with the data collected via a complex sampling design. In particular, household survey data are collected via stratified two-stage unequal probability cluster sampling, which requires special care when modeled. 
\end{abstract}



\end{frontmatter}

\section{Introduction}

There has been an increased interest in using covariates for spatial data modeling \citep{uwiringiyimana2022bayesian, macharia2019sub}, with previous studies showing that the inclusion of influential spatial covariates can lead to improved prediction accuracy \citep{lindstrom2014flexible, ZERAATPISHEH2022105723}. It is straightforward to include covariates in a linear model, within the linear predictor of spatial random effects models, but the extension to more flexible covariate models is much more difficult. Flexible modeling is desirable to leverage nonlinearities and interactions which can lead to improved predictive performance.


Currently, Gaussian random field (GRF) models are commonly used as a modeling framework for capturing spatial correlations, with wide application, for example, in population health modeling \citep{diggle2019model}. {\color{black}For data that arise from complex survey designs, fully Bayesian modeling approaches have been explored, see for example \cite{chan2020bayesian, banerjee2024finite} and \cite{finley2024models}.} A common approach to computation in spatial modeling is Integrated nested Laplace approximation (INLA) \citep{rue2009approximate}, particularly in a low-and-middle-income countries (LMIC) context \citep{utazi2018high, burstein2019mapping}. INLA is a powerful tool for carrying out Bayesian inference, but requires the predictors to have a linear form (though this does include spline modeling, since such models can be expressed in linear form), which fails to capture multivariate nonlinearities and interactions and may lead to reduced predictive performance.

On the other hand, there are many machine learning methods that allow for flexible covariate modeling, including the stacked generalization method \citep{davies2016optimal} and random forests \citep{ren2018} and these have been applied to spatial modeling problems \citep{osgood2018mapping, shi2021digital}. However, these approaches suffer from drawbacks. {\color{black}A number of machine learning approaches in spatial modeling ignore the spatial dependence \citep{georganos2021geographical} while others (including the aforementioned references) do not correctly propagate uncertainty as is required for valid inference \citep{daw2023reds,wakefield2019estimating}}. 

Bayesian Additive Regression Trees (BART) \citep{chipman2010bart} provide reliable Bayesian inference by specifying a prior distribution on the `sum-of-trees' structure that flexibly models the covariates. Previous applications of BART on spatial data modeling problems include \cite{muller2007spatially}, who proposed a spatial BART model based on a conditional autoregressive (CAR) model, and \cite{Krueger2020}, who suggested using a matrix exponential spatial specification (MESS) \citep{lesage07} as an extension of CAR.  \cite{tan2019bayesian} review extensions to the basic BART  model, including a description of the approach due to \cite{muller2007spatially}.  However, compared to the GRF that we use, these models specify a simple variance-covariance structures and are designed for area-level data. The model considered by \cite{spanbauer2021nonparametric} assumes a more generic covariance structure under a non-spatial setting, but is limited to models with few random effects and does not scale well to large spatial datasets. As a result, we note that the large numbers of random effects in continuous spatial models, along with the strong dependence in the posterior, lead to computational difficulties and pose serious challenges for incorporating BART into spatial models.  

In this paper, we propose \textbf{BARTSIMP}, which is shorthand for `\textbf{BART} for \textbf{S}patial \textbf{I}NLA \textbf{M}odeling and \textbf{P}rediction', as a GRF spatial Bayesian modeling framework with a flexible covariate regression model. Our model leverages the flexibility in BART to capture the nonlinearity and interactions across covariates, while also recognizing the complex spatial correlation structure in the residuals which is modeled by the GRF. To ease computation, we use the INLA-within-MCMC method \citep{gomez2018markov} to design a Metropolis-within-Gibbs sampler that integrates out the random effects using INLA and then performs MCMC updates on the remaining parameters. {\color{black} This model is the first to simultaneously model the covariate effects through BART and the spatial effects through a GRF, while producing Bayesian uncertainty estimates (credible intervals). Our spatial BART model is also the first to use INLA as an approximate approach to reduce the computational burden.}

We organize the paper as follows. We will describe our motivating example, which is spatial prediction for child anthropometric data, in Section \ref{sec:mot}. The model formulation is in Section \ref{sec:md} and we describe the Metropolis-within-Gibbs sampler which we use for model implementation in Section \ref{sec:comp}. We apply our model to simulated data experiments in Section \ref{sec:sim} and return to the Kenya data in Section \ref{sec:app}. Section \ref{sec:conc} concludes the paper with a discussion.

\section{Motivating Dataset}
\label{sec:mot}

Our study is motivated by child undernutrition data collected in the 2014 Kenya Demographic and Health Survey (DHS). Specifically, we are interested in wasting for children under the age of five, which is measured using the weight-for-height Z-score (WHZ) metric. The Z-score can be interpreted as the number of units the weight for a child is higher or lower than the median value, compared to all children of the same height in the population \citep{mei2007standard}. For example, wasting is defined by a WHZ score below $-2$ \citep{kassie2019exploring}. 

The data is collected via stratified two-stage cluster sampling. In the first stage, 1584 enumeration areas (EAs) were sampled across 92 strata. The 92 sampling strata are defined according to the 47 counties and the urban-rural status of the county (with Nairobi and Mombasa county  only having urban areas). In the second stage, 40,300 households were sampled across the selected EAs, yielding 20,977 individual WHZ measurements. 

We use geospatial covariates on the raster level, with a list of covariate descriptions in Table \ref{tab:cov}. In our example, for simplicity, we did not include the urban-rural status as a covariate in our model, which may lead to some bias due to oversampling of urban clusters but we believe that the inclusion of population density provides some protection.

\begin{table}[]
\small
\centering
\caption{DHS Geospatial covariates used in our model.\\}
\begin{tabular}{lll}
\hline
\textbf{Variable} &
  \textbf{Years} &
  \textbf{Source} \\ \hline
Population Density &
  2014 &
  \begin{tabular}[c]{@{}c@{}}WorldPop\tablefootnote{http://www.worldpop.org.uk/data/get\_data/.} \end{tabular} \\ \hline
Night Time Light &
  2013 &
  \begin{tabular}[l]{@{}l@{}}National Centers for\\ Environmental Information\tablefootnote{https://ngdc.noaa.gov/eog/dmsp/downloadV4composites.html.} \end{tabular} \\ \hline
Vegetation Index &
  2013-2014 &
  \begin{tabular}[l]{@{}l@{}}NASA EOSDIS Land Processes\\ DAAC\tablefootnote{https://lpdaac.usgs.gov/dataset\_discovery/modis/modis\_products\_table/mod13a3\_v006.}\end{tabular} \\ 
\hline
Average Temperature &
  \begin{tabular}[l]{@{}l@{}}Mean value over\\ the period\\ 1970--2000\end{tabular} &
  \begin{tabular}[c]{@{}c@{}}WorldClim\tablefootnote{http://worldclim.org/version2.}\end{tabular} \\ \hline 
Precipitation &
  \begin{tabular}[l]{@{}l@{}}Mean value over\\ the period\\ 1970--2000\end{tabular} &
  \begin{tabular}[c]{@{}l@{}}WorldClim\tablefootnote{http://worldclim.org/version2.}\end{tabular} \\ \hline
Access to Nearest City &
  2000 &
  \begin{tabular}[l]{@{}l@{}}Joint Research Centre of the\\ European Commission\tablefootnote{http://forobs.jrc.ec.europa.eu/products/gam/download.php.}\end{tabular} \\ \hline
\end{tabular}
\label{tab:cov}
\end{table}

\begin{figure}[]
\centering
\includegraphics[width = \linewidth]{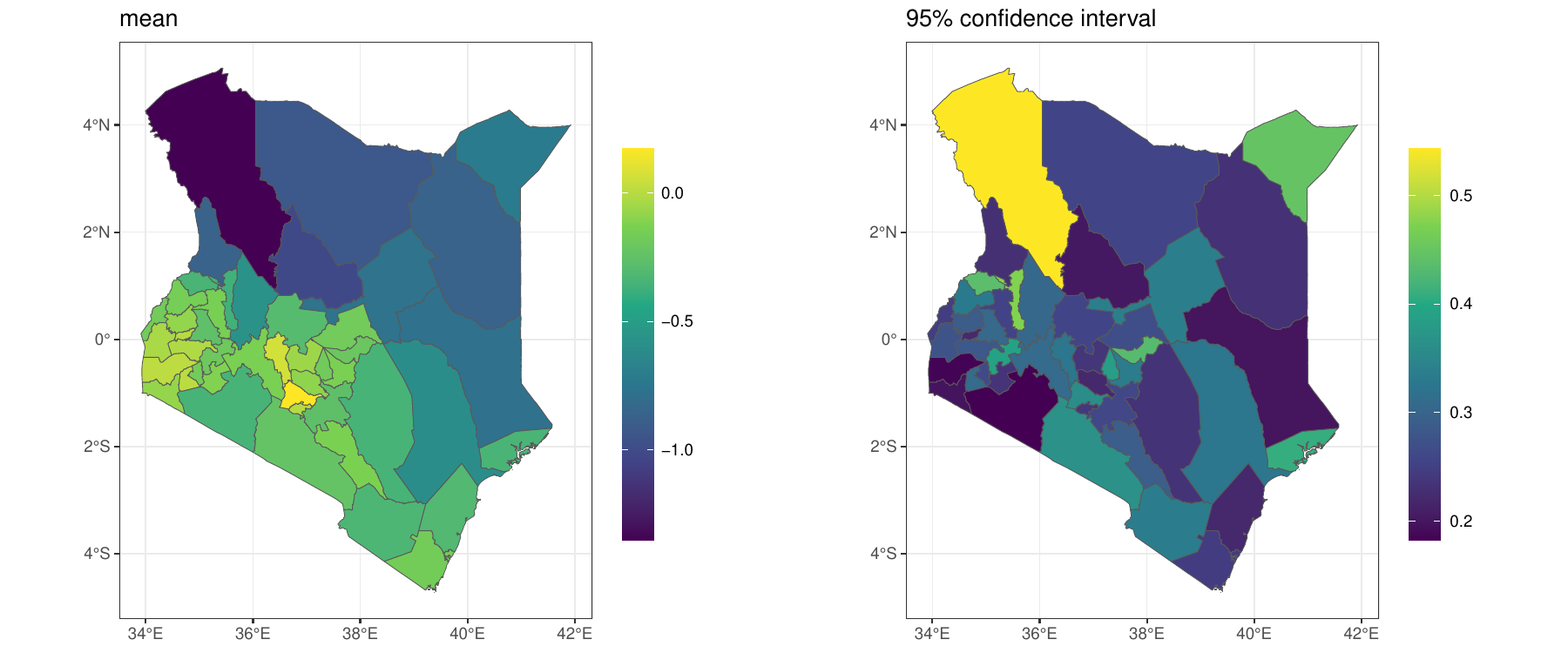}
\caption{Maps of Admin 1 level WHZ mean estimates (left) and 90\% confidence interval widths (right).}
\label{fig:direct}
\end{figure}

The aim of the analysis is to provide predictions of WHZ at various administrative levels. To introduce some terminology, Admin 0 denotes the national level, Admin 1 one below that (for example, states in the United States) and Admin 2 one below that (for example, counties in the United States). In Kenya, the Admin 1 level contains 47 counties, while Admin 2 contains 290 constituencies. Hence, our objective is a problem in small area estimation (SAE). If there are sufficient data in each area, a weighted (so-called direct) estimator can be used. But often there are insufficient area-based data and models must be introduced. In an area-based model \citep{fay1979estimates} the direct estimator is modelled and random effects are introduced. In a unit-level model, point level data are modeled \citep{battese1988error}, and this is the path we follow. \cite{rao2015small} provides a comprehensive overview of SAE. For the Kenya data, and at Admin 1, we plot direct estimators and the width of a 90\% confidence interval in Figure \ref{fig:direct}. In the left panel, we see  higher levels of WHZ in the west and south, with lower, less desirable, outcomes in the north and east. In the right panel, we see that the accompanying uncertainty measures are wide, so that the direct estimates are producing imprecise estimates. There are higher levels, in particular, in the northwest and it is hard to make definitive statements given the large uncertainty in this part of the map. Hence, we aim to reduce uncertainty using spatial smoothing and covariate modeling. We emphasize that we model at the point level, but the final deliverable for policy making is at the area-level, so that we are required to aggregate from points to areas.


\section{Model Description}
\label{sec:md}
To describe the data framework and introduce notation, we will describe a conventional traditional spatial model with a linear model in Section \ref{sec:LGM}. We will then introduce BART in Section \ref{sec:BART} and propose our spatial version in Section \ref{sec:SBART}. We describe the priors in Section \ref{sec:SBARTprior}.

\subsection{Model formulation and preliminaries}
\label{sec:LGM}



We consider the spatial dataset $(\mathbf{Y}, \boldsymbol{x})$, where $\mathbf{Y}$ is the collection of observed outcomes and $\boldsymbol{x}$ is the covariate matrix. Let $n$ be the number of spatial locations observed (in our example this corresponds to clusters), noting that we allow multiple outcomes (these are children in our case) to be observed at the same location so that $\boldsymbol{y} = \left(\boldsymbol{y}_1, ..., \boldsymbol{y}_n\right)^\intercal$ is a jagged array of vectors, with $n_i$ being the number of observations in cluster $i$ and $\boldsymbol{y}_i = \left(y_{i1}, \dots, y_{in_i}\right)^\intercal$ being the vector of all observations at the same location. Specifically, we let $y_{ik}:=y_{k}(\boldsymbol{s}_i)$ be the $k$-th response observed at location $\boldsymbol{s}_i$, $i = 1, \dots, n$. We let $P$ be the number of covariates (including the intercept), so that $\boldsymbol{x}_i = (x_{i1}, \dots, x_{iP})$ is the vector of covariates observed at location $\boldsymbol{s}_i$ where $x_{ip}$ is the value of covariate $p$ observed at cluster $i$ for $p = 1, \dots, P$ (for simplicity of notation, we assume that observations at the same spatial unit share the same covariate values). We assume a Gaussian error model for the observations with measurement error  variance $\sigma_e^2$. We let the latent random spatial component at location $\boldsymbol{s}_i$ be denoted ${z}(\boldsymbol{s}_i)$, $i=1,\dots,n$, and assumed to follow a Gaussian random field (GRF) with Mat\'ern covariance function over the region. 

We will describe a flexible covariate model in Section \ref{sec:BART}, but we begin by describing a conventional linear model.
A linear mixed effects model with a spatial random field can be written as: 
{\color{black}
\begin{align} 
y_{ik} \mid \boldsymbol{x}_{i}, \boldsymbol{\beta},  \boldsymbol{z}, \sigma_e^2 &\stackrel{i.i.d}{\sim} N\left(\boldsymbol{x}_i^{\top}\boldsymbol{\beta} +z_i, \sigma_e^2\right), \label{eq:model0} \\ 
\boldsymbol{z} \mid \boldsymbol{\psi}  &\sim G F(0, \sigma_m^2  \boldsymbol{\Sigma}),
\label{eq:model}
\end{align}
$k=1,\dots,n_i$, $i=1,\dots,n$.
The  
$n \times n$ matrix $\mathbf{\Sigma}$ is a Mat\'ern correlation matrix and $\sigma_m^2$ is the spatial variance. The correlation matrix has elements,
\begin{equation} 	
\Sigma_{i j} =\operatorname{Corr}_M\left(z\left(\boldsymbol{s}_i\right), z\left(\boldsymbol{s}_j\right)\right)=\frac{2^{1-\nu} }{\Gamma(\nu)}\left(\kappa\left\|\boldsymbol{s}_i-\boldsymbol{s}_j\right\|\right)^\nu K_\nu\left(\kappa\left\|\boldsymbol{s}_i-\boldsymbol{s}_j\right\|\right),
\label{eq:matern}
\end{equation} 
for $i,j=1,\dots,n$, $i \neq j$, and 
with $\kappa$ being the scale parameter of the GRF. Note that the smoothness parameter $\nu$ is usually fixed \textit{a priori}, because data sparsity makes it difficult to estimate. There is a one-to-one relationship between the scale parameter $\kappa$ and the range parameter (which is the more commonly interpreted parameter) with $\rho=\frac{\sqrt{ 8 \nu}}{\kappa}$. We let $\boldsymbol{\theta} = (\sigma_e^2, \boldsymbol{\psi})$ with $\boldsymbol{\psi} = (\sigma_m^2, \rho)$ being the set of Mat\'ern covariance parameters.}  The model is completed with priors on the regression parameters $\boldsymbol{\beta}$ and a prior on the variance components,
\begin{equation}\label{eq:varprior}
\pi( \sigma_e^2,\sigma_m^2,\kappa).
\end{equation}
Details on the particular prior we use are postponed until Section \ref{sec:SBARTprior}.

For latent Gaussian models (LGMs), a closed-form expression for $\pi(\boldsymbol{z}, \boldsymbol{\beta}, \boldsymbol{\theta} \mid \boldsymbol{y})$ is computationally expensive for large $n$, due to the computation of the inverse and determinant of the dense $n \times n$ covariance matrix. Efficient MCMC sampling algorithms are difficult for spatial models because of the strong dependence between parameters \citep{knorr2002block}. \cite{rue2009approximate} showed that the INLA method, an approximation Bayesian inference procedure based on Laplacian approximations and numerical integration rules for LGMs is available for models where the latent Gaussian random field is distributed as a Gaussian Markov random field (GMRF). The INLA approach provides approximations for the marginal distributions of the latent Gaussian random field and the hyperparameters, along with a reliable estimate for the marginal distribution $\pi(\boldsymbol{y})$ \citep{hubin2016estimating}.

For models with continuously indexed Gaussian fields, \cite{lindgren2011explicit} showed that a discrete GMRF can be used to approximate the continuously indexed data, with a GRF generated through a Mat\'ern covariance function, via a stochastic partial differential equation (SPDE) approach. This not only reduces the computation complexity from $\mathcal{O}(n^{3})$ to $\mathcal{O}(n^{3/2})$, but also enables model inference to be carried out through INLA. 

\subsection{The BART model}
\label{sec:BART}

The Bayesian Additive Regression Trees (BART) model \citep{chipman2010bart} can be viewed as the summation of a fixed number of highly flexible nonparametric regression functions, where each function {\color{black}$g(\cdot ; T_j, \boldsymbol{\mu}_j): \mathbb{R}^p \rightarrow \mathbb{R}, ~ j = 1, \dots, m$} is a random function that maps the $P$-dimensional covariate space onto the real line, with a total number of $l = 1, \dots, m$ functions. The component functions are defined by two sets of parameters, the binary tree structure $\mathbf{T}=\left(T_1, \ldots, T_m\right)$ and the terminal node values $\boldsymbol{\mu} = (\boldsymbol{\mu}_1, \ldots, \boldsymbol{\mu}_m)$. Each binary tree can be viewed as a set of decision rules that splits the covariate space into a finite number of regions: at any internal nodes for a given tree, a `splitting variable' is chosen from the $P$ covariates and a threshold value is chosen that splits the current region into two sub-regions, and a fitted value is assigned to the corresponding region at each end node, according to the terminal node values $\boldsymbol{\mu}$. By splitting according to different covariate variables and threshold values, BART is capable of modeling complex nonlinear relationships and interactions among covariates. Furthermore, a regularization prior that penalizes tree complexity in order to avoid model overfitting is assigned to the tree structure parameters so that the trees functions are `weak learners' \citep{chipman2010bart}. 
\subsection{Sampling Model and Latent Field}
\label{sec:SBART}
The sampling model for BARTSIMP is a combination of the linear mixed effects model described in Section \ref{sec:LGM} and the BART terms in Section \ref{sec:BART}. We substitute the linear covariate effects for the sum-of-tree model to obtain: 
\begin{align*}
	y_{ik} \mid z_i, \boldsymbol{x}_{i p}, \mathbf{T}, \boldsymbol{\mu}, \sigma_e^2 \stackrel{i . i . d}{\sim} N\left(\sum_{l=1}^m g\left(\boldsymbol{x}_{i p} ; T_l, \boldsymbol{\mu}_l\right)+z_i, \sigma_e^2\right)
\end{align*}
with the distribution of $\boldsymbol{z}$ and Mat\'ern covariance as previously defined in Equations \eqref{eq:model} and \eqref{eq:matern}.


\subsection{The Priors}
\label{sec:SBARTprior}
We assume a priori independence between the tree parameters $(\mathbf{T}, \boldsymbol{\mu})$, the residual variance parameter $\sigma_e^2$ and the spatial hyperparameters $\boldsymbol{\psi}$:
\begin{align*}
	p\left(\mathbf{T}, \boldsymbol{\mu}, \sigma_e^2, \boldsymbol{\psi}\right)=\left[\prod_{j=1}^m p\left(\boldsymbol{\mu}_j \mid T_j\right) p\left(T_j\right)\right] p\left(\sigma_e^2\right) p\left (\boldsymbol{\psi}\right).
\end{align*}
{\color{black} For the tree parameters and the residual variance parameter, we follow the prior specifications in \cite{chipman1998bayesian} and decompose the tree structure prior into three hierarchical parts: the probability of a node being non-terminal, the probability of choosing one of the covariates as the splitting variable for a non-terminal node and the probability of choosing a splitting value given the chosen splitting variable. For a tree node at depth $d$, the probability of it being a non-terminal node is $\alpha(1+d)^{-\beta}$, with $\alpha \in(0,1), \beta \in[0, \infty)$. In the  examples in the paper, we defer to the default settings in \cite{chipman2010bart} and choose $\alpha = 0.95$ and $\beta = 2$. Further, we assume that the splitting variable is uniformly chosen among all covariates for each internal node, and the splitting value is uniformly chosen from all distinct values of the selected splitting variable.} We assume an independent and identically distributed normal prior for the terminal node values $\boldsymbol{\mu}_1, \dots, \boldsymbol{\mu}_m$ given their corresponding tree structure $T_j$. The mean and variance for the normal prior are chosen such that each tree will only function as a `weak learner' that contributes a small part in the model; we refer to \cite{chipman2010bart} for further details. 

For the residual variance parameter $\sigma_e^2$, we take the default choice in \cite{chipman2010bart} and specify a scaled inverse-Gamma distribution $\sigma_e^2 \sim \nu \lambda / \chi_\nu^2$, where $\nu$ is the degrees-of-freedom. The values of the hyperparameters $(\nu, \lambda)$ are chosen through an exploratory data analysis procedure to give $\hat{\sigma}_r^2$ as an empirical guess of $\sigma_e^2$ through a tentative working model (for example, an SPDE model with all covariates as linear predictors), which matches the $q$-th quantile of the scaled inverse Gamma prior. Following \cite{chipman2010bart}, we let $(q, \nu) = (0.9, 3)$. Hence, the prior depends on the data in a weak fashion. Finally, we place a penalized complexity (PC) prior on the Mat\'ern hyperparameters $\boldsymbol{\psi}$, following \cite{Fuglstad19}. The joint PC prior for $\boldsymbol{\psi}$ is 
\begin{align*}
	p(\boldsymbol{\psi}) = \frac{d}{2}\tilde{\lambda}_1\tilde{\lambda}_2\rho^{-d/2 - 1} \exp \left(-\tilde{\lambda}_1 \rho^{-d/2} - \tilde{\lambda}_2 \sigma_m \right), \sigma_m >0, \rho > 0,
\end{align*}
where $d=2$, $\tilde{\lambda}_1 = -\log(\alpha_1)\rho_0^{d/2}$ and $\tilde{\lambda}_2 = -{\log (\alpha_2)}/{\sigma_0}$. The hyperparameters $(\alpha_1, \alpha_2, \rho_0, \sigma_0)$ guarantee that 
$\operatorname{P}(\rho < \rho_0) = \alpha_1$ and $\operatorname{P}(\sigma_m > \sigma_0) = \alpha_2$. We let $\alpha_1 = \alpha_2 = 0.5$ and set $(\rho_0, \sigma_m)$ to the crude estimates obtained from fitting a model with only an intercept and the spatial random field. Hence, we again specify a data dependent prior.

\section{Algorithm}
\label{sec:comp}

Under our model framework, a standard MCMC algorithm requires the sampling and storage of the spatial random field $\boldsymbol{z}$, which is computationally undesirable. Hence, our strategy is to integrate out $\boldsymbol{z}$ and operate on the corresponding marginal likelihood. However, the calculation of the marginal likelihood involves computing the determinant of large and dense variance-covariance matrices, which becomes prohibitive as the number of datapoints at distinct locations (clusters in the Kenya example) becomes large. As a result, we consider the `INLA-within-MCMC' technique proposed by \cite{gomez2018markov}. Specifically, the algorithm calls INLA to approximate the aforementioned marginal likelihood, which is then used to compute the Metropolis-Hastings acceptance ratio in the MCMC routine for the remaining model parameters. 

{
\color{black}
We begin this section with a brief overview of the INLA-within-MCMC method. We let $\boldsymbol{\theta}$ denote the whole ensemble of hyperparameters and latent effects. Further, we consider the following partition $\boldsymbol{\theta} = \left(\boldsymbol{\theta}_{c}, \boldsymbol{\theta}_{-c}\right)$, where $\boldsymbol{\theta}_{-c}$ represents the set of parameters we would like to integrate out, and $\boldsymbol{\theta}_{c}$ being the set of parameters for which we will use a Metropolis-Hastings algorithm. 
The posterior distribution can be re-expressed as 
$$
\pi(\boldsymbol{\theta} \mid \boldsymbol{y}) \propto \pi\left(\boldsymbol{y} \mid \boldsymbol{\theta}_{-c}, \boldsymbol{\theta}_{c}\right) \pi\left(\boldsymbol{\theta}_{-c} \mid \boldsymbol{\theta}_{c}\right) \pi\left(\boldsymbol{\theta}_{c}\right) .
$$
Integrating $\boldsymbol{\theta}_{-c}$ from both sides gives,
\begin{align}
\label{eq:int_pos}
  \pi\left(\boldsymbol{\theta}_{c} \mid y\right) \propto \pi\left(\boldsymbol{y} \mid \boldsymbol{\theta}_{c}\right) \pi\left(\boldsymbol{\theta}_{c}\right),  
\end{align}
In the context of the Metropolis-Hastings algorithm, let $\boldsymbol{\theta}_c^*$ be the potential value for the new iteration proposed by the transition kernel $q\left(\cdot\mid \boldsymbol{\theta_c}\right)$. A new $\boldsymbol{\theta}_c$ is proposed within each iteration of the MCMC routine. Plugging \eqref{eq:int_pos} into the expression of the acceptance probability, we have, 
\begin{align}
\label{eq:acceptance_ratio}
\alpha=\min \left\{1, \frac{\pi\left(\boldsymbol{y} \mid \boldsymbol{\theta}_{c}^{*}\right) \pi\left(\boldsymbol{\theta}_{c}^{*}\right) q\left(\boldsymbol{\theta}_{c} \mid \boldsymbol{\theta}_{c}^{*}\right)}{\pi\left(\boldsymbol{y} \mid \boldsymbol{\theta}_{c}\right) \pi\left(\boldsymbol{\theta}_{c}\right) q\left(\boldsymbol{\theta}_{c}^{*} \mid \boldsymbol{\theta}_{c}\right)}\right\}.
\end{align}

With the prior distribution $\pi(\cdot)$ 
and proposal distribution $q(\cdot \mid \cdot)$ for $\boldsymbol{\theta}_c$ specified, the computation bottleneck in \eqref{eq:acceptance_ratio} is to calculate $\pi\left(\boldsymbol{y} \mid \boldsymbol{\theta}_{c}\right)$ and $\pi\left(\boldsymbol{y} \mid \boldsymbol{\theta}_{c}^*\right)$. In practice, it can be computationally expensive to obtain an exact likelihood. The INLA-within-MCMC approach provides an efficient method to provide approximations to $\pi\left(\boldsymbol{y} \mid \boldsymbol{\theta}_{c}\right)$ and $\pi\left(\boldsymbol{y} \mid \boldsymbol{\theta}_{c}^*\right)$, by fitting models with R-INLA, with the values of $\boldsymbol{\theta_c}$ (or $\boldsymbol{\theta_c}^*$) fixed.

We next provide an overview of the Metropolis-within-Gibbs algorithm we will be using for the BARTSIMP model, where the INLA-within-MCMC approach is applied to approximate the acceptance ratio. } We let $T_{-j}, \boldsymbol{\mu}_{-j}$  be the set of trees and terminal node values, respectively, not including $T_j$ and $\boldsymbol{\mu}_j$. An outline of the MCMC routine is given in Algorithm \ref{algo:bartsimp}. We now describe each of the updates.
\begin{algorithm}[!htb]
\SetAlgoLined
\textbf{Input:} Initialized values for $\mathbf{T}, \boldsymbol{\mu}, \sigma_e^2, \psi$ for a number of MCMC iterations $B$. \\
\For{$b \gets 1$ \KwTo $B$}{
\For{$j \gets 1$ \KwTo $m$}{
  \textbf{Update} $T_j \mid \boldsymbol{y}, T_{-j}, \boldsymbol{\mu}_{-j}, \sigma_e^2, \boldsymbol{\psi}$. \\ 
  \textbf{Update} $\boldsymbol{\mu}_j \mid \boldsymbol{y}, T_j, T_{-j}, \boldsymbol{\mu}_{-j}, \sigma_e^2, \boldsymbol{\psi}$. \\ 
  \textbf{Update} $\sigma_e^2 \mid \boldsymbol{y}, \mathbf{T}, \boldsymbol{\mu}, \boldsymbol{\psi}$. \\ 
  \textbf{Update} $\boldsymbol{\psi} \mid \boldsymbol{y}, \mathbf{T}, \boldsymbol{\mu}, \sigma_e^2$. \\ 
  }
}
 \caption{An overview of the BARTSIMP algorithm.}
 \label{algo:bartsimp}
\end{algorithm}\\

\subsection{Update \texorpdfstring{$T_j \mid \boldsymbol{y}, T_{-j}, \boldsymbol{\mu}_{-j}, \sigma_e^2, \psi$}{u1}}
\label{sec:treeupdate}
To update the sum-of-trees variables, we follow the Bayesian backfitting MCMC algorithm that is outlined in Section 3.1 of \cite{chipman1998bayesian}. Specifically, we carry out a sequential update of the set of tree structure parameters and terminal node values $\{(T_1, \boldsymbol{\mu}_1), \dots, (T_m, \boldsymbol{\mu}_m)\}$, updating each tree, one at a time. To update each of the tree structure parameters $T_j$, we  integrate out $\boldsymbol{\mu}_j$ and do a Metropolis-within-Gibbs update of $T_j$, while keeping the other parameters, $(T_{-j}, \boldsymbol{\mu}_{-j}, \sigma_e^2, \boldsymbol{\psi})$, fixed. The acceptance ratio for $T_j$ is,
\begin{align}
\label{eq:T_accept}
	\alpha_{T_j} = \min \left\{1, \frac{\pi\left(\boldsymbol{y} \mid T_{j}^{*}, T_{-j}, \boldsymbol{\mu}_{-j}, \sigma_e^2,\boldsymbol{\psi}\right) \pi\left(T_{j}^{*}\right) q\left(T_{j}\mid T_{j}^{*}\right)}{\pi\left(\boldsymbol{y} \mid T_{j}, T_{-j}, \boldsymbol{\mu}_{-j}, \sigma_e^2, \boldsymbol{\psi}\right) \pi\left(T_{j}\right) q\left(T_{j}^{*} \mid T_{j}\right)}\right\}.
\end{align}
Here, $T_j^*$ is the proposed structure for tree $j$, following the proposal distribution for $T_j$ in \cite{chipman2010bart}, through which we can compute the transition kernel ratio $
q\left(T_j \mid T_j^*\right)
/
 q\left(T_j^* \mid T_j\right)
$. To compute the likelihood ratio, we employ the backfitting algorithm as follows. For the likelihood $\pi\left(\boldsymbol{y} \mid T_{j}, T_{-j}, \boldsymbol{\mu}_{-j}, \sigma_e^2, \boldsymbol{\psi}\right)$, we see that, given the tree parameters of all trees except for $T_j$, we can compute the residual values with respect to the remaining $m-1$ trees. Denote these vectors of residuals as $\boldsymbol{r}^{(j)}_{1}, \dots, \boldsymbol{r}^{(j)}_{n}$, where 
$${r}_{ik}^{(j)} = {y}_{ik}- \sum_{l \neq j} g\left(\boldsymbol{x}_{i} ; T_{l}, \boldsymbol{\mu}_{l}\right), \qquad l = 1,\dots,m, ~~ i = 1, \dots ,n, ~~ k = 1, \dots, n_i.$$
We see that the original model is now equivalent to a single-treed  model (along with the spatial field and the Gaussian noise) with response $(\boldsymbol{r}_1^{(j)}, \dots, \boldsymbol{r}_{n}^{(j)})$, where the superscript $(j)$ is with respect to the particular tree that is being `left out'. Since the tree structure for $T_j$ is conditioned upon, the tree part is equivalent to a linear model ${\mathbf{C}^{(j)}}\boldsymbol{\mu}_j$ where $\boldsymbol{\mu}_j = (\mu_{j1}, ..., \mu_{jb_j})$ is the set of terminal node parameters in $\boldsymbol{\mu}_j$, and ${\mathbf{C}^{(j)}}$ is the $n \times b_j$ covariate matrix with elements,
$$C^{(j)}_{it} =  1 \text{ if } \boldsymbol{x}_{[i]} \text{ belongs to terminal node } t \text{ in tree } T_j, \text{ otherwise } C^{(j)}_{it} = 0,$$ 
where $C^{(j)}_{it}$ is the element in the $i$-th row and $t$-th column of ${\mathbf{C}^{(j)}}$. Thus, given $T_j, T_{-j}, \boldsymbol{\mu}_{-j}$, the whole model is equivalent to:
{
\color{black}
\begin{equation} 
\begin{aligned}
\boldsymbol{r}^{(j)} \mid {\mathbf{C}^{(j)}} , \boldsymbol{\mu}_j, \sigma_e^2, \boldsymbol{\psi} &\sim N\left({\mathbf{C}^{(j)}} \boldsymbol{\mu}_j, \sigma_e^2\mathbf{I}_{n} + \mathbf{\Sigma}(\boldsymbol{\psi}) \right) \\
	\boldsymbol{\mu}_j  &\sim N(\boldsymbol{0}, \kappa\mathbf{I}_{b_j}).
\end{aligned}
\label{eq:samp}.
\end{equation} 
Therefore, the likelihood term $\pi\left(\boldsymbol{y} \mid T_j, T_{-j}, \boldsymbol{\mu}_{-j}, \sigma_e^2, \boldsymbol{\psi}\right)$ in \eqref{eq:T_accept} can be expressed as 
\begin{equation}
\begin{aligned}
\label{eq:marg_lik}
    \pi\left(\boldsymbol{y} \mid T_j, T_{-j}, \boldsymbol{\mu}_{-j}, \sigma_e^2, \boldsymbol{\psi}\right) &= \pi(\boldsymbol{r}^{(j)} \mid {\mathbf{C}^{(j)}} (\boldsymbol{x}), \sigma_e^2, \boldsymbol{\psi}) \\&= \mathcal{N}(\mathbf{0}; \mathbf{0}, \kappa {\mathbf{C}^{(j)}} {\mathbf{C}^{(j)}}^\intercal + \mathbf{\Sigma} + \sigma_e^2 \mathbf{I}_n),
\end{aligned}
\end{equation}
where $\mathcal{N}(\boldsymbol{y} ; \boldsymbol{\mu}, \mathbf{\Sigma})$ denotes the likelihood for a Gaussian distribution with mean $\boldsymbol{\mu}$ and covariance matrix $\mathbf{\Sigma}$, evaluated at $\boldsymbol{y}$. On examination of \eqref{eq:marg_lik}, we see that a major computational challenge for generating posterior samples for $\boldsymbol{\mu}_j$ arises from computing the inverse of $\left(\kappa {\mathbf{C}^{(j)}} {\mathbf{C}^{(j)}}^\intercal + \mathbf{\Sigma} + \sigma_e^2 \mathbf{I}_n\right)$, which has complexity $\mathcal{O}(n^3)$.
To reduce the computational burden, we first re-express the first line in \eqref{eq:samp} as a hierarchical model, with the spatial effect $\boldsymbol{z}$ being conditioned upon:
\begin{equation} 
\begin{aligned}
\boldsymbol{r}^{(j)} \mid {\mathbf{C}^{(j)}}, \boldsymbol{z}, \sigma_e^2 &\stackrel{i . i . d}{\sim} N\left({\mathbf{C}^{(j)}} \boldsymbol{\mu}_j+\boldsymbol{z}, \sigma_e^2\mathbf{I}_{d} \right) \\
	\boldsymbol{z} \mid \boldsymbol{\psi} &\sim GF(\boldsymbol{0}, \boldsymbol{\Sigma}(\boldsymbol{\psi})).
\end{aligned}
\label{eq:samp2}
\end{equation} 
The GRF in \eqref{eq:samp} can be approximated as a GMRF \citep{lindgren2011explicit}:
\begin{equation} 
\begin{aligned}
\boldsymbol{r}^{(j)} \mid {\mathbf{C}^{(j)}}, \mathbf{A}, \boldsymbol{u}, \sigma_e^2 &\stackrel{i . i . d}{\sim} N\left({\mathbf{C}^{(j)}} \boldsymbol{\mu}_j+\mathbf{A}\boldsymbol{u}, \sigma_e^2\mathbf{I}_{d} \right) \\
	\boldsymbol{u}&\sim N(\boldsymbol{0}, \mathbf{Q}^{-1})
\end{aligned}
\label{eq:samp3}
\end{equation}
where $\boldsymbol{u}$ is the GMRF, $\mathbf{A}$ is a projection matrix that projects the random effects on mesh points to the spatial locations, and $\mathbf{Q}$ is the sparse precision matrix for $\boldsymbol{u}$. Thus we can approximate the likelihood term in \eqref{eq:marg_lik} as $\mathcal{N}\left(\mathbf{0} ; \mathbf{0}, \kappa {\mathbf{C}^{(j)}} {\mathbf{C}^{(j)}}^{\top}+\mathbf{A}\mathbf{Q}^{-1}\mathbf{A}^\intercal+\sigma_e^2 \mathbf{I}_n\right)$. The computation for $\pi\left(\boldsymbol{y} \mid T_j^*, T_{-j}, \boldsymbol{\mu}_{-j}, \sigma_e^2, \boldsymbol{\psi}\right)$ is similar.
\subsection{Update \texorpdfstring{$\boldsymbol{\mu}_j \mid \boldsymbol{y}, T_{j}, T_{-j}, \boldsymbol{\mu}_{-j}, \sigma_e^2, \psi$}{u2}}
The model formulation in \eqref{eq:samp} leads to the posterior distribution for $\boldsymbol{\mu}_j$: 
\begin{align*}
	\boldsymbol{\mu}_j \mid {\mathbf{C}^{(j)}}, \boldsymbol{r}^{(j)}, \sigma_e^2, \boldsymbol{\psi}, \kappa \sim N(\boldsymbol{m}_j, \mathbf{V}_j)
\end{align*}
where 
\begin{equation}
\begin{aligned}
\label{eq:postsamp_mu}
	\mathbf{V}_j &= \left[{\mathbf{C}^{(j)}}^\intercal \left(\sigma_e^2 \mathbf{I}_n + \mathbf{\Sigma}\right)^{-1}{\mathbf{C}^{(j)}} + \kappa^{-1}\mathbf{I}_{b_j}\right]^{-1} \\ \boldsymbol{m}_j &= \mathbf{V}_j^{-1} {\mathbf{C}^{(j)}}^\intercal \left(\sigma_e^2 \mathbf{I}_n + \mathbf{\Sigma}\right)^{-1} \boldsymbol{r}^{(j)}.
\end{aligned}
\end{equation}
Here, the major computation burden in \eqref{eq:postsamp_mu} is to compute the inverse of $\left(\sigma_e^2 \mathbf{I}_n + \mathbf{\Sigma}\right)$. Similar to Section \ref{sec:treeupdate}, we can approximate the GP with GMRF and expresse the approximated posterior distribution for $\boldsymbol{\mu}_j$ as
\begin{align*}
	\boldsymbol{\mu}_j \mid {\mathbf{C}^{(j)}}, \boldsymbol{r}^{(j)}, \sigma_e^2, \boldsymbol{\psi}, \kappa \sim N(\tilde{\boldsymbol{m}}_j, \tilde{\mathbf{V}}_j)
\end{align*}
where 
\begin{equation} 
\begin{aligned}
\label{eq:postsamp_mu_approx}
	\tilde{\mathbf{V}}_j &= \left[{\mathbf{C}^{(j)}}^\intercal \left(\sigma_e^2 \mathbf{I}_n + \mathbf{AQ}^{-1}\mathbf{A}\right)^{-1}{\mathbf{C}^{(j)}} + \kappa^{-1}\mathbf{I}_{b_j}\right]^{-1} \\ \tilde{\boldsymbol{m}}_j &= \mathbf{V}_j^{-1} {\mathbf{C}^{(j)}}^\intercal \left(\sigma_e^2 \mathbf{I}_n + \mathbf{AQ}^{-1}\mathbf{A}\right)^{-1} \boldsymbol{r}^{(j)}.
\end{aligned}
\end{equation}
To further simplify the computation, note that the most computationally costly steps in computing \eqref{eq:postsamp_mu_approx} are inverting $\mathbf{Q}$ and $\sigma_e^2 \mathbf{I}_n + \mathbf{AQ}^{-1}\mathbf{A}$. By the Woodbury matrix identity we have,
\begin{align}
\label{eq:woodbury}
	\left(\sigma_e^2 \mathbf{I}_n + \mathbf{AQ}^{-1}\mathbf{A}\right)^{-1} &= \sigma_e^{-2}\mathbf{I} - \sigma_e^{-4}\mathbf{A}(\mathbf{Q} + \sigma_e^{-2}\mathbf{A}^\intercal\mathbf{A})^{-1} \mathbf{A}^\intercal.
\end{align}
Note that compared to \eqref{eq:postsamp_mu_approx}, in this expression, we do not need to explicity invert $\mathbf{Q}$, and also note that $\mathbf{Q} + \sigma_e^{-2}\mathbf{A}^\intercal\mathbf{A}$ is a sparse matrix, which leads to much lower computational cost.
}

\subsection{Update \texorpdfstring{$\sigma_e^2 \mid \boldsymbol{y}, \mathbf{T}, \boldsymbol{\mu}, \boldsymbol{\psi}$}{u3} and \texorpdfstring{$\boldsymbol{\psi} \mid \boldsymbol{y}, \mathbf{T}, \boldsymbol{\mu}, \sigma_e^2$}{u4}}
As with the tree structure parameters $(T_1, \dots, T_m)$, we can use the Metropolis-within-Gibbs method to update the residual variance parameter $\sigma_e^2$ and use the INLA-within-MCMC technique to approximate the Metropolis-Hastings acceptance ratio:
$$
\alpha_{\sigma_e^2}=\min \left\{1, \frac{\pi\left(\boldsymbol{y} \mid \mathbf{T}, \boldsymbol{\mu}, \psi, \sigma_e^{2 *}\right) \pi\left(\sigma_e^{2 *}\right)q\left(\sigma_e^{2} \mid \sigma_e^{2 *}\right)}{\pi\left(\boldsymbol{y} \mid \mathbf{T}, \boldsymbol{\mu}, \psi, \sigma_e^{2}\right) \pi\left(\sigma_e^{2}\right)q\left(\sigma_e^{2 *} \mid \sigma_e^{2}\right)} \right\}.
$$
To ensure that the variance parameter is positive, we use a Gaussian proposal for $\log \sigma_e^2$. We again use the backfitting technique to compute the overall residual $\boldsymbol{r}$, defined as all $m$ tree terms subtracted from the response value:
$$
r_{ik}=y_{ik}-\sum_{l = 1}^m g\left(\boldsymbol{x}_i ; \mathbf{T}_l, \boldsymbol{\mu}_l\right), \qquad k=1,\ldots, n_i,~~ i=1, \ldots, n.
$$
This is equivalent to the following sampling model:
$$
\begin{gathered}
\boldsymbol{r} \mid \boldsymbol{z}, \sigma_e^2 \stackrel{i . i . d}{\sim} N\left(\boldsymbol{z}, \sigma_e^2\right) \\
\boldsymbol{z} \mid \boldsymbol{\psi} \sim G F(\boldsymbol{0}, \boldsymbol{\Sigma}).
\end{gathered}
$$
By fitting a linear mixed effect model with residual variance fixed at $\sigma_e^2$ and spatial hyperparameters fixed at $\boldsymbol{\psi}$, we can compute the approximated marginal likelihood of the model $\pi(\boldsymbol{r} \mid \sigma_e^2, \boldsymbol{\psi})$, which is equivalent to $\pi\left(\boldsymbol{y} \mid \mathbf{T}, \boldsymbol{\mu}, \boldsymbol{\psi}, \sigma_e^{2}\right)$. We update the spatial hyperparameters in a similar fashion. 

\subsection{R and C++ Integration}

We implemented the backbone of the MCMC algorithm for BARTSIMP based on the C++ code in the \texttt{BART} package \citep{spanbauer2021nonparametric}. In order to use the functions in the \texttt{R-INLA} package \citep{martins2013bayesian} to carry out the INLA computation during the Metropolis-Hastings adjustment step, we used Rcpp \citep{eddelbuettel2011rcpp, eddelbuettel2013seamless, eddelbuettel2018extending} as an interface between C++ and R. To provide an open-source computing software for BARTSIMP, we developed the R package \texttt{BARTSIMP}, with source code available at \url{https://github.com/AlexJiang1125/BARTSIMP}. 

\section{Simulation Experiments}
\label{sec:sim}

\subsection{Simulation setting}

In this section, we study several simulation scenarios in which the spatial and covariate signals have different strengths. We consider a $50 \times 50$ grid surface over a study region $[0,1] \times [0,1]$ and denote the set of grid cells as $G$. For the covariates associated with each grid cell, we independently generate two covariates from a uniform distribution on $[0,1]$. Let $x_{gp}, g \in G, p = 1,2$, represents the $p$-th covariate for grid cell $g$ and let $\boldsymbol{x}_g=(x_{g1},x_{g2})$. The actual deterministic field evaluated at cell $g$, denoted $f(\boldsymbol{x}_g)$, is defined as:
\begin{align*}
	f(\boldsymbol{x}_g) = (1-\omega) \boldsymbol{z}_g^* + \omega f_{0}(\boldsymbol{x}_g), ~~ \forall g \in G,
\end{align*}
where the `baseline' spatial field $\boldsymbol{z}^*$ is generated from a GRF with Mat\'ern parameters $\kappa = 2.5$, $\sigma_m^2 = 0.5$. Likewise, we let $f_{0}(\boldsymbol{x}_g)$ be the `baseline' covariate surface generated based on the {\color{black} function:
\begin{align*}
    f_{0}(\boldsymbol{x}_g) = \begin{cases}
    3 & x_{g1} < 0.5 \\
    0 & x_{g1}\geq 0.5, x_{g2} \geq 0.5 \\
    -2 & x_{g1}\geq 0.5, x_{g2} < 0.5,
    \end{cases}
\end{align*}
Finally, the scalar $\omega$ is fixed at different values and can be interpreted as the proportion of `covariate signal' among the overall signal. Here we consider five different scenarios for $\omega$: $1$ (covariate signal only), $0.8$ (strong covariate signal), $0.5$ (medium covariate signal), $0.2$ (weak covariate signal) and $0$ (spatial signal only), which we denote as scenarios 1, 2, 3, 4 and 5.} We then randomly select 250 cells from the grid and randomly sample one location uniformly within each cell, ending up with 250 spatial points over the study region, mimicking the spatial locations for the clusters. The number of observations for each spatial location is sampled from a uniform distribution over $\{5,6,7,8,9,10\}$. For each observation at a given spatial location, its value is defined as the sum of the deterministic field value (spatial field plus covariate signal)  from the grid cell it belongs to, and Gaussian random noise with $\sigma_e^2 = 1$. We simulate $10$ datasets for each scenario. For the BARTSIMP and BART model, we used {\color{black}ensembles of 20 trees, and collected 2,000 posterior samples after a burn-in period of 2,000.} 

\subsection{Performance Criteria}
\label{sec:perfc}
We compare our model against the following three alternatives: a standard BART model without a spatial latent field \textbf{(BART)}, a GMRF spatial model with Mat\'ern covariance function fitted by the SPDE approach, including all covariates in a linear model \textbf{(SPDE)} and a spatial SPDE model, with intercept only, i.e.,~no covariates, \textbf{(SPDE0)}. {\color{black}Additionally, we implemented a modified version of BARTSIMP, in which  the likelihood function is calculated using the exact formula, instead of the SPDE approximation \textbf{(BARTSIMP-EXACT)}. We include this approach to assess whether the INLA approximation to the marginal distribution is accurate.

We examine the models via performance measures that evaluate both point and interval estimates, over the gridded surface:} 
\begin{itemize}
        {\color{black}\item \textbf{Root mean squared error (RMSE)}: Let $\hat{f}^j(\boldsymbol{x}_g)$ be the model prediction for $f(\boldsymbol{x}_g)$ given by method $j$, the RMSE is defined as:
        \begin{align*}
            \text{RMSE}^j = \frac{1}{|G|}\sum_{g \in G}\left({\hat f^j(\boldsymbol{x}_g}) - f(\boldsymbol{x}_g)\right)^2, \qquad j =1,\ldots, 5.
        \end{align*}
        \item \textbf{Average interval length (AIL)}: Let $(L_{g,\alpha}^{j}, U_{g,\alpha}^{j})$ denote the $100 \times (1 - \alpha) \%$ prediction interval for $f(\boldsymbol{x}_g)$, available from method $j$. The AIL is defined as the average length of prediction intervals over the gridded surface:
        \begin{align*}
            \text{AIL}^j = \frac{1}{|G|}\sum_{g \in G}|U_{g, \alpha}^j - L_{g, \alpha}^j|, \qquad j =1,\ldots, 5
        \end{align*}}
	\item \textbf{Average coverage rate (ACR)}: 
    The ACR is defined as the coverage of $f(\boldsymbol{x}_g)$, for the interval, $(L_{g,\alpha}^{j}, U_{g,\alpha}^{j})$:
	\begin{align*}
		\text{ACR}^{j} = \frac{1}{|G|} \sum_{g \in G} \operatorname{I}\left(f(\boldsymbol{x}_g) \in L_{g,\alpha}^{j}, U_{g,\alpha}^{j})\right), \qquad j =1,\ldots, 5.
	\end{align*}
	For AIL and ACR we take $\alpha = 0.05$.
	\item \textbf{Average Interval score (AIS, \citealp{gneiting2007strictly})}: the AIS is an integrated metric for prediction interval accuracy defined as: 
	\begin{multline*}
		\text{AIS}^j = \frac{1}{|G|} \sum_{g \in G} \left[ (U_{g,\alpha}^{j} -  L_{g,\alpha}^{j}) + \frac{2}{\alpha} \left(L_{g,\alpha}^{j} - f(\boldsymbol{x}_g) \right) \cdot \mathbf{1}(f(\boldsymbol{x}_g) < L_{g,\alpha}^{j})\right. + \\ \left.
		\frac{2}{\alpha} \left(f(\boldsymbol{x}_g) - U_{g,\alpha}^{j}\right) \cdot \mathbf{1}(f(\boldsymbol{x}_g) > U_{g,\alpha}^{j}) \right],  \qquad j =1,\ldots, 5.
	\end{multline*}
The AIS can be broken down into three parts, with the first part penalizing  wider intervals and the second and third parts penalizing low coverage rates. Among the measures, RMSE assesses the point prediction accuracy, while AIL, ACR and AIS focus on the uncertainty prediction accuracy. 
	
\end{itemize}

\subsection{Results}

\begin{figure}
    \centering
    \includegraphics[width=\linewidth]{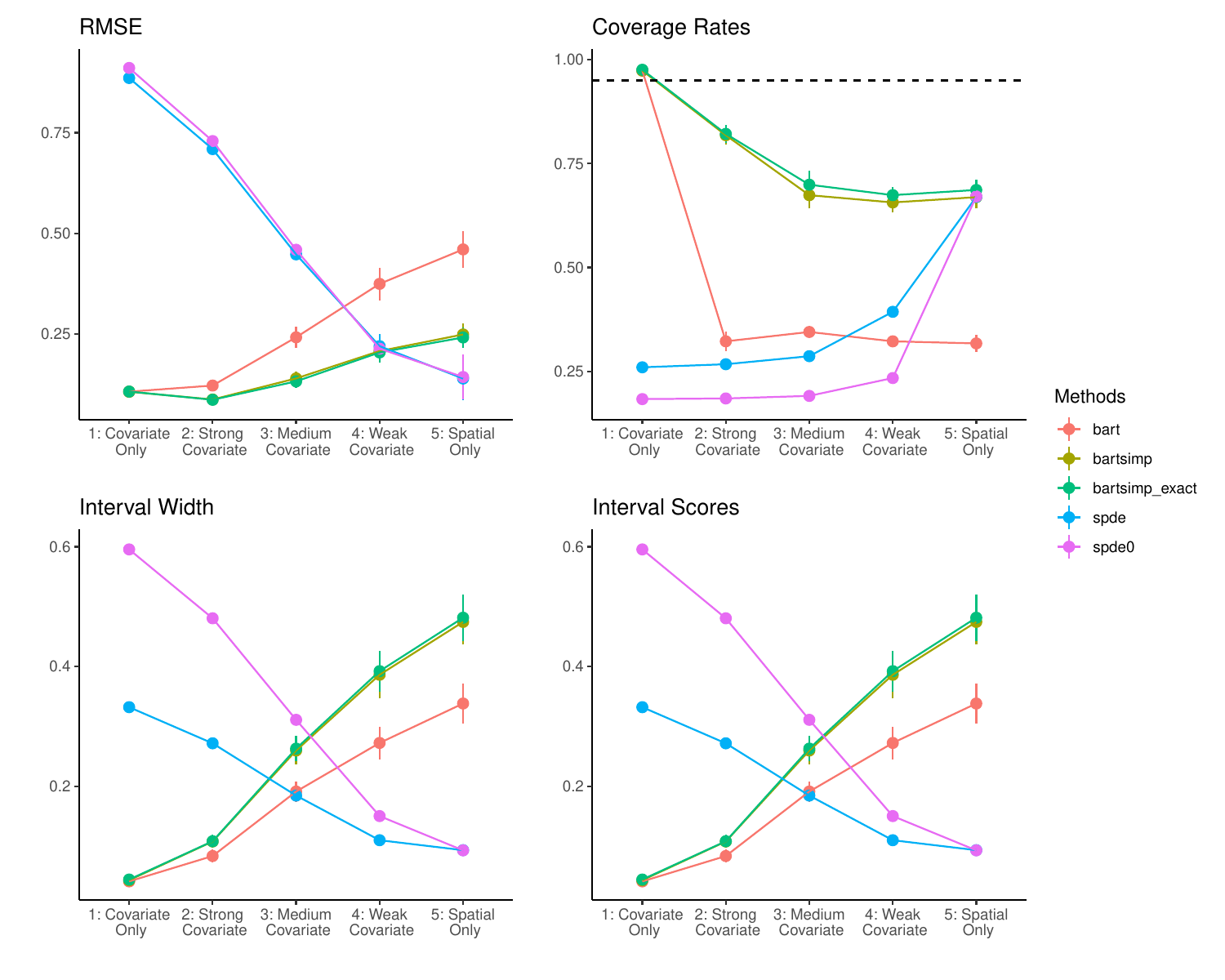}
    \caption{Root mean squared error (RMSE), average coverage rate (ACR), average interval length (AIL) and average interval score (AIS) for five different methods under five scenarios (covariate only, strong, medium, weak covariate signals and spatial signals only). The dots show the average values over 10 replications, and the vertical lines illustrate the 95\% Wald type interval computed from the standard deviation over 10 replications. The horizontal dashed line in the upper right figure illustrates the 95\% nominal coverage.}
    \label{fig:modcomp}
\end{figure}

{\color{black}Figure \ref{fig:modcomp} shows comparisons of RMSE, AIL, ACR and AIS across all five methods under the five scenarios. First, we observe that BARTSIMP and BARTSIMP-EXACT perform similarly for all four metrics, suggesting that the approximation error from SPDE is negligible in model fitting. For point prediction accuracy, BART (the covariate-only model) has the lowest RMSE when there is only covariate signal, and both SPDE and SPDE0 (the spatial models) have the lowest RMSE when there is only spatial signal. Meanwhile, BARTSIMP (and BARTSIMP-EXACT) perform the best across all methods for the remaining scenarios where the true data is generated from a mixture of covariate and spatial signals. For uncertainty estimation, BARTSIMP performs best out of all methods in all scenarios, having coverage rates closest to the nominal coverage. As spatial signals become stronger, SPDE and SPDE0 tend to have lower interval widths and interval scores, and both BART and BARTSIMP tend to have wider interval widths and interval scores.

BARTSIMP and BARTSIMP-EXACT perform similarly, but their computation time differs. A scalability test that compares the run times of evaluating a Gaussian log-likelihood using the exact likelihood and our approximation is conducted in Section 2 of the Supplementary Materials. The experiment shows that while the exact method is quicker for data with few observations, the approximation method is considerably more efficient as the number of observations grows. Attaning the coverage is an issue for all approaches, particular for the stronger spatial signal cases, but we think this is just the reality of modeling point-level spatial data without large sample sizes. The same behavior of under-coverage was seen in \cite{osgood2023statistical}. This was a comprehensive examination and Figure 2 of this paper in particular clearly shows this phenomenon. 

Finally, a sensitivity analysis on choices for the model hyperparameters is conducted in Section 3 of the Supplementary Materials, where the performance metrics are analyzed for the BARTSIMP method, under 25 datasets simulated from scenario 2 with difference values of $(\alpha_1, \alpha_2, \sigma_0, \rho_0, \nu, q)$. The results show that there are no clear differences in model performance, under choices of model hyperparameters within reasonable ranges. 
}

\section{Application}
\label{sec:app}

In this section, we investigate the prediction performance of BARTSIMP, as compared to BART, SPDE and SPDE0, evaluated on the WHZ measurements from the 2014 Kenya DHS. 

\subsection{Cross-Validation Exercise}

To compare the prediction performances across different approacahes, we apply cross-validation and split the 1584 clusters in the 2014 DHS survey into training and test datasets of cluster sizes 1267 and 317, respectively. To preserve the stratification structure of the design in both data sets, we consider stratified sampling of the clusters such that the strata proportions in the training set roughly matches that in the test set. Finally, we repeated the procedure 10 times to reduce sampling variation caused by using a single data split. {\color{black}We again use RMSE, ACR, AIL and AIS as performance criteria for all four methods, similar to Section \ref{sec:perfc}, and we use the same settings as before for the BART and BARTSIMP models. The algorithm took one day to run for BARTSIMP on Ubuntu 18.04, with 50GB of memory.

Table \ref{fig:methcomp} shows ACR (nominal coverage is 95\%), RMSE, AIL and AIS for all approaches. The results are the average over 10 test datasets, with standard deviations in brackets. We see that BARTSIMP and BART dominate the other two methods in terms of having clostest to nominal coverage.  The BARTSIMP model has the lowest AIS.} 
\begin{table}[t]
\centering
\caption{Prediction performance measures over all four competing methods. The nominal coverage is 95\% and small values of AIS are preferred. The averages over $10$ test datasets are shown with standard deviations shown in brackets.\\}
\begin{tabular}{rlrrrrr}
  \hline
 & Model & ACR & RMSE & AIL & AIS \\ 
  \hline
1 & BART-SIMP & \textbf{98}\% (0.003) & 0.584 (0.024) & 6.882 (0.082) & \textbf{0.172} (0.002) \\
  2 & BART & \textbf{93}\% (0.005) & 0.590 (0.024) & 5.287 (0.028) & 0.182 (0.005)  \\ 
  3 & SPDE & 78\% (0.010) & \textbf{0.565} (0.023) & \textbf{3.136} (0.030) & 0.256 (0.009)\\ 
  4 & SPDE0 & 78\% (0.008) & 0.578 (0.025) & 3.156 (0.032) & 0.257 (0.010) \\ 
   \hline
\end{tabular}
\label{fig:methcomp}
\end{table}
\subsection{Point and Areal Prediction}

In this section, we conduct spatial predictions on a grid surface over the study region and generate aggregated estimates at the Admin 1 and Admin 2 levels.
 The reason we produce estimates at the areal-level is that resource allocation and policy making decisions are made at the area-level.

At location $\boldsymbol{s}$, we let $\text{WHZ}(\boldsymbol{s})$ be the spatial surface of the height-for-weight Z-scores and $d_{5}(\boldsymbol{s})$ be the under-five population density. The areal level WHZ is a weighted average over the under-five population density $d_5(\boldsymbol{s})$, as the Z-scores were evaluated for children under age five. The  $d_5(\boldsymbol{s})$ values are obtained from WorldPop (\url{https://www.worldpop.org}). The WHZ for an administrative region $R_i$ is 
\begin{align}
\label{eq:sp}
	\text{WHZ}_{R_i} = \frac{\int_{R_i} \text{WHZ}(\boldsymbol{s}) d_5(\boldsymbol{s})}{\int_{R_i}  d_5(\boldsymbol{s})}, \qquad \qquad i = 1,2, \dots, m,
\end{align}
where $m$ is the number of administrative areas.

\begin{figure}[!ht]
\centering
\includegraphics[width = .85\linewidth]{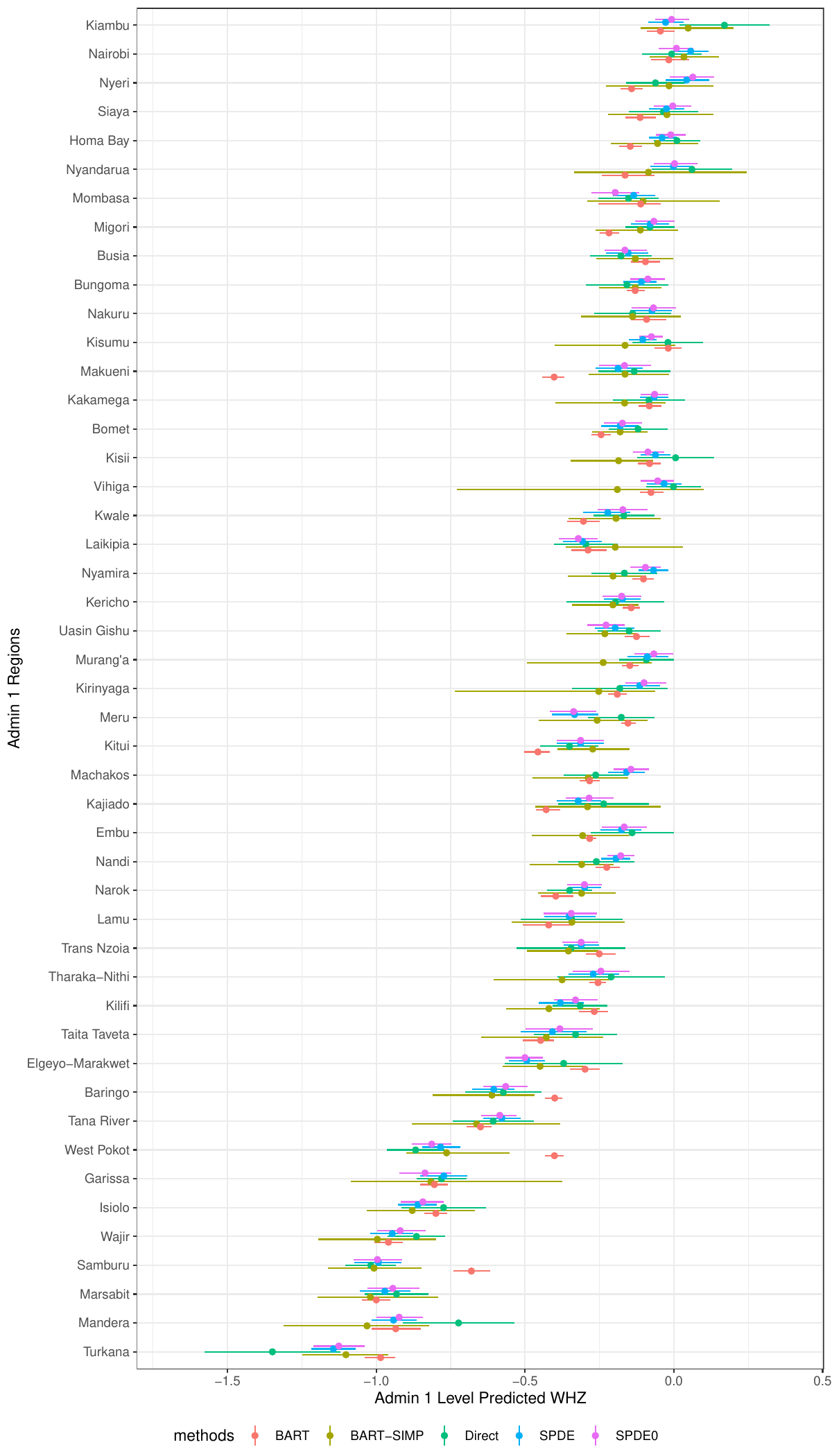}
\caption{The Admin 1 level posterior median and 95\% credible/confidence intervals of WHZ for BART, BARTSIMP, SPDE0, SPDE and direct (weighted) estimates. The Admin 1 areas are arranged according to the predicted posterior mean given by BARTSIMP.}
\label{fig:admin1comps}
\end{figure}

\begin{figure}[!ht]
\centering
\includegraphics[width = .49\linewidth]{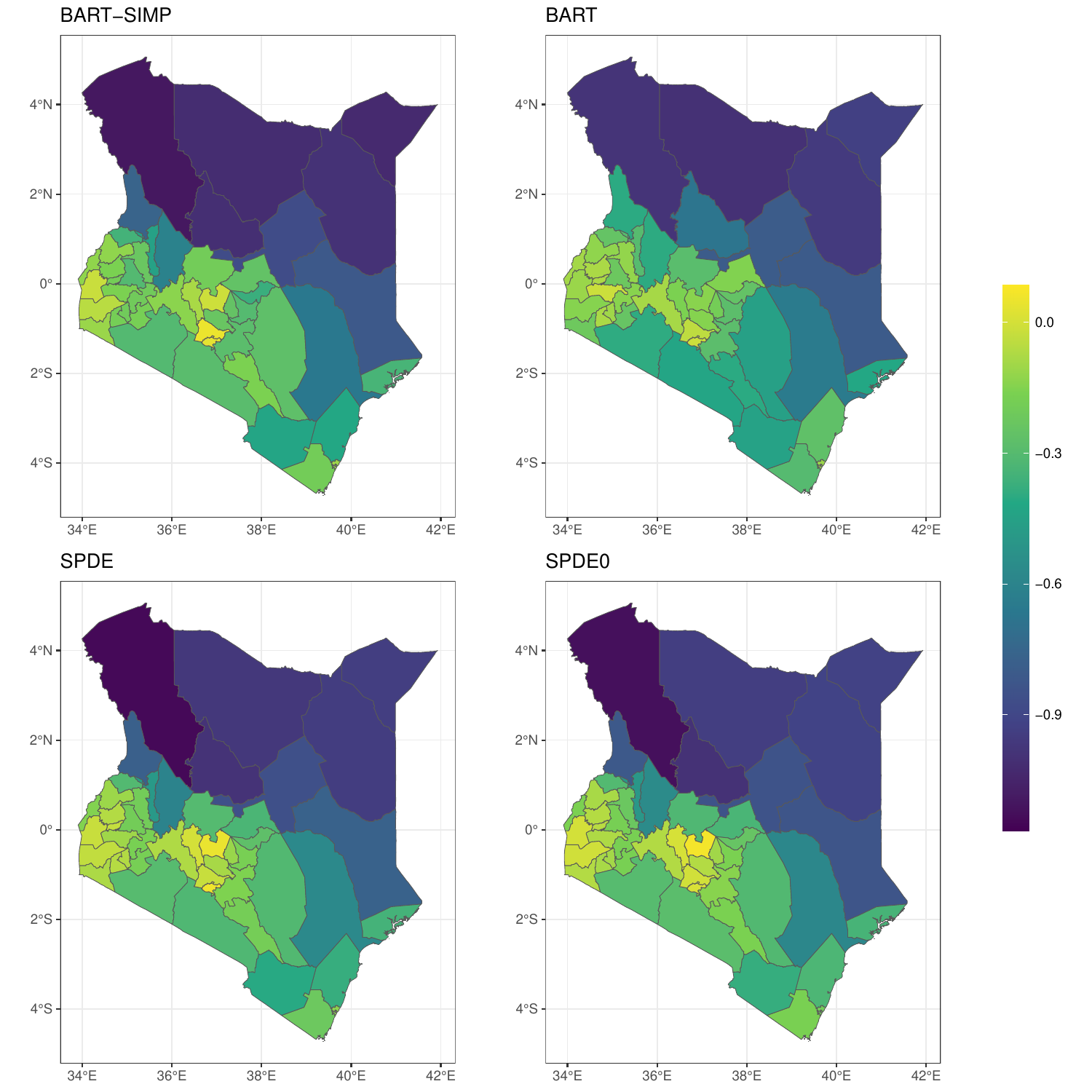}
\includegraphics[width = .49\linewidth]{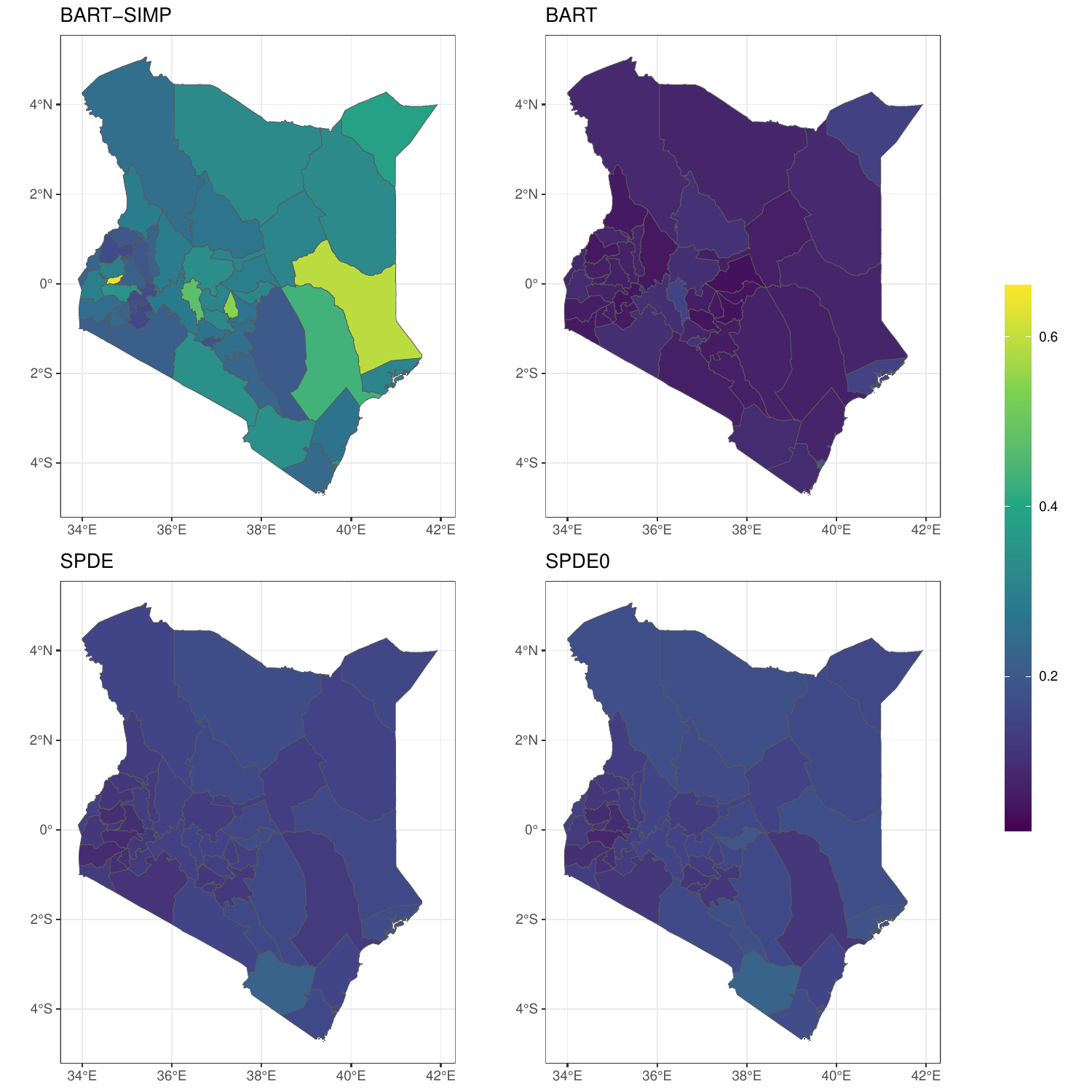}
\caption{Maps of Admin 1 level WHZ posterior median (left) and 90\% credible interval lengths (right) for BARTSIMP, BART, SPDE and SPDE0.}
\label{fig:admin1areas}
\end{figure}

We approximate the integrals in \eqref{eq:sp} by a weighted sum over observations on grid cells located at $\boldsymbol{s}_g$ over the regions, $g \in R_i$. Let $\text{WHZ}(\boldsymbol{s}_g)$ be the height-for-weight Z-score and $d_5(\boldsymbol{s}_g)$ be the under-five population density evaluated at grid cell $g$. The regional WHZ, approximated on the grid, is
\begin{align}
\label{eq:WHZ}
	{\text{WHZ}}_{R_i} \approx  \frac{\sum_{g \in R_i} {\text{WHZ}}(\boldsymbol{s}_g) d_5(\boldsymbol{s}_g) }{\sum_{g \in R_i} d_5(\boldsymbol{s}_g)}.
\end{align}
Using \eqref{eq:WHZ}, we can calculate the posterior mean and $95\%$ credible interval quantiles for regional ${\text{WHZ}}$ at all Admin 1 and Admin 2 areas, based on the four methods. As a comparison, we also consider the direct weighted areal-level estimates and 95\% (design-based) confidence intervals. Figure \ref{fig:admin1comps} shows the posterior median and 95\% credible/confidence intervals, derived from all five methods, and Figure \ref{fig:admin1areas} shows the predicted areal-level WHZ for all 47 Admin 1 areas in Kenya. The predicted posterior mean for WHZ given by BARTSIMP ranges from -1.10 to 0.05 over Admin 1 regions, showing that there is large within country variation in WHZ in Kenya. Among the 47 Admin 1 regions, Kiambu and Nairobi have the highest WHZ predictions -- these are the most populated counties in Kenya. Low WHZ scores occurred in areas such as Turkana, Mandera and Marsabit and these could be targeted for interventions. 

We see from Figures \ref{fig:admin1comps} and \ref{fig:admin1areas} that while all five methods give quantitatively similar overall patterns of the areal estimate across the regions, BARTSIMP and the spatial methods provide similar point estimates at the local level. BARTSIMP gives relatively wider credible interval lengths, which  is consistent with the simulations, in which we saw that  these intervals are more appropriate. Finally, BART does not yield reliable point estimates, and gives interval estimates that are far too narrow. 

We also provided areal-level predictions on Admin 2 levels, with results shown in the Supplemental Materials. Examination of these results show that the direct estimates for Admin 2 regions have more unreliable point estimates with much wider confidence intervals, due to insufficient samples observed in each region. 

\subsection{Partial dependence}

\begin{figure*}[!ht]
	\centering
	\includegraphics[width = \linewidth]{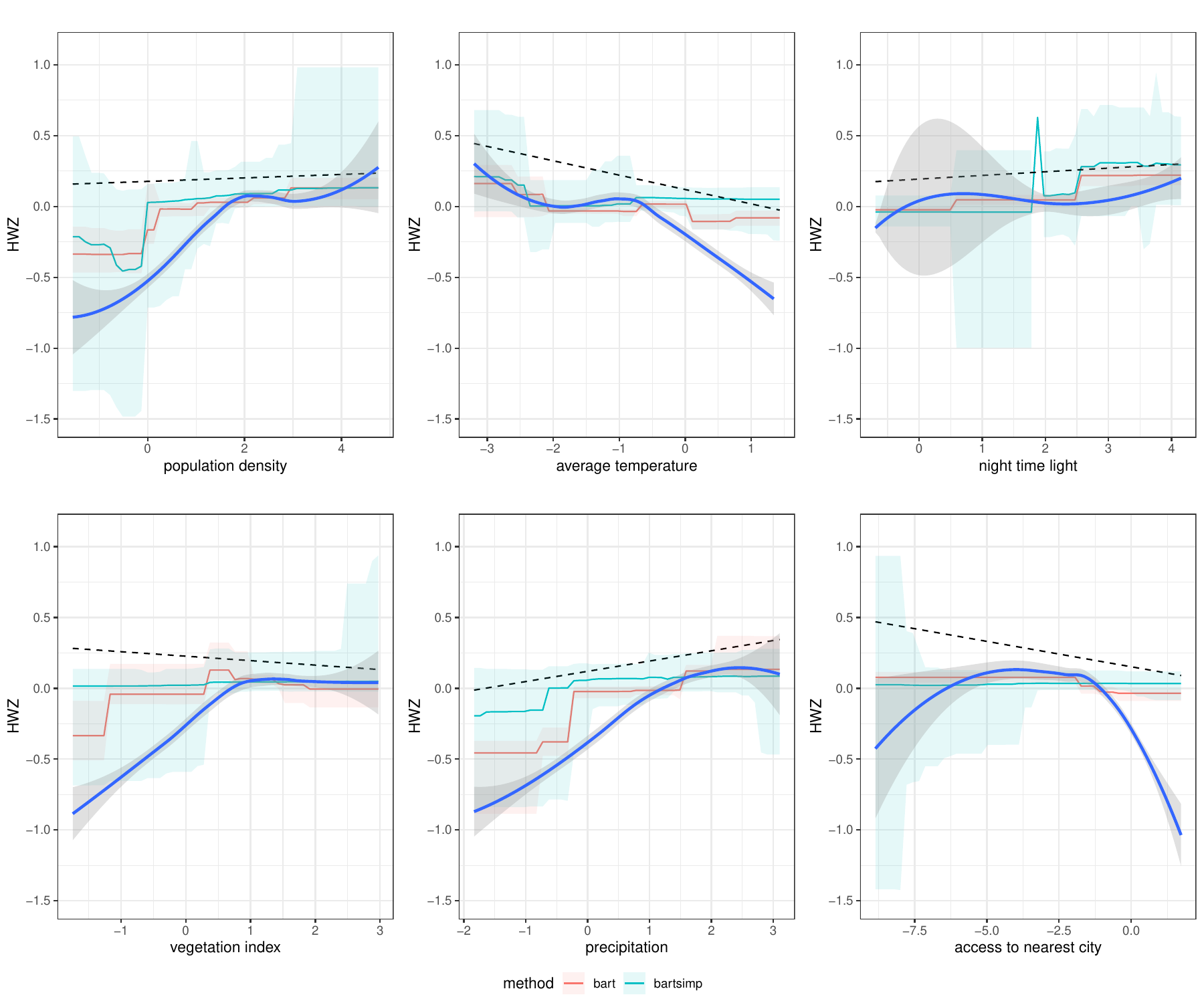}
	\caption{Partial dependence function for all six covariates (population density, average temperature, night time light, vegetation index, precipitation and access to nearest city) and the 95\% pointwise credible interval provided by the BARTSIMP (blue) and BART (red) methods. The blue lines and grey bands represent the locally-weighted smoother functions and their 95\% pointwise confidence interval. }
	\label{fig:pd1}
\end{figure*}

In our example, it is also of  interest to study the marginal influence of each variable. Various measures of partial dependence have been proposed in the machine learning literature \citep{breiman2001random, goldstein2015peeking}.  For models based on BART, the partial dependence function \citep{friedman2001greedy} is a commonly used measure. Let $f(\cdot): \mathbb{R}^p \rightarrow \mathbb{R}$ be a multivariate function defined on $p$ variables $\boldsymbol{x}= (x_1, \dots, x_p) $. Let $\boldsymbol{x}_t$ denote the of variables we wish to study, out of the $p$ variables, and let $\boldsymbol{x}_c = \boldsymbol{x} \slash {x}_t$ be their compliment. Suppose we have $n$ observations of such multivariate variables. The partial dependence function for $f(\cdot)$ with respect to ${x}_t$ is defined as 
\begin{align*}
	f^{pd}({x}_t) = \frac{1}{n} \sum_{i=1}^n f({x}_t, \boldsymbol{x}_{ic}),
\end{align*}
where $\boldsymbol{x}_{i.}, i = 1, \dots, n$ represents the $i$-th observation. For BARTSIMP and BART, we can calculate the partial dependence function of the sum-of-trees function for all six variables in our dataset. Figure \ref{fig:pd1} shows the partial dependence function and its $95\%$ credible interval based on 1000 MCMC samples for the BARTSIMP and BART methods. As a comparison, we also included the locally-weighted smoother function estimated from the scatterplot of centered WHZ versus each variable. We observe that for all six variables, BARTSIMP and BART have similar partial dependence patterns that roughly resemble the pattern of the raw data. Substantively, holding all other variables constant, WHZ increases with increasing population density, vegetation index and precipitation, and decreases with increasing average temperature. The association with access is nonlinear but there is great uncertainty at low levels of access. Note that BART has much narrower credible intervals, which support our previous finding that BART tends to underestimate model uncertainty in this setting. 

\section{Discussion}
\label{sec:conc}

In this work, we have proposed BARTSIMP as a novel framework for flexible covariate modeling and prediction for spatial datasets. We incorporated the nonparametric nature of BART into continuous spatial models and leveraged the flexibility of BART to allow nonlinear covariate relationships and interactions. We have developed a sampling-based inference algorithm for BARTSIMP based on the INLA-within-MCMC technique. 

BARTSIMP has a number of limitations which require more investigation. A cross-validation analysis showed that the BARTSIMP method yields average coverage rates closer to the nominal coverage compared to other methods while having poorer estimation performance than other methods when the covariate signal is weak. This suggests that when the covariate signal in the dataset is not strong compared to the spatial signal then it is not worth attempting any flexible covariate modeling. This is supported by our simulation studies. We hope our study provides inspiration for future spatial modeling studies with complex covariate patterns. Another potential extension is to consider non-Gaussian likelihood models where the outcome variable is discrete (i.e.,~counts and proportions) using the P\'{o}lyâ Gamma data augmentation technique \citep{polson2013bayesian}.

The key challenge that arises when one combines  spatial models with machine learning techniques, is attaching an appropriate measure of uncertainty. In general, uncertainty estimation is difficult with machine learning techniques, since the bootstrap does not work in many instances. For example, with sparse estimators this occurs because the limiting distribution is complex and may not be continuous, see \citealp{dezeure2015high}. When combining machine learning techniques with spatial models, this aspect becomes even more challenging. 

\bibliographystyle{apalike}
\bibliography{main}

\begin{thebibliography}{}

\bibitem[Battese et~al., 1988]{battese1988error}
Battese, G.~E., Harter, R.~M., and Fuller, W.~A. (1988).
\newblock An error-components model for prediction of county crop areas using survey and satellite data.
\newblock {\em Journal of the American Statistical Association}, 83(401):28--36.

\bibitem[Breiman, 2001]{breiman2001random}
Breiman, L. (2001).
\newblock Random forests.
\newblock {\em Machine Learning}, 45(1):5--32.

\bibitem[Burstein et~al., 2019]{burstein2019mapping}
Burstein, R., Henry, N.~J., Collison, M.~L., Marczak, L.~B., Sligar, A., Watson, S., Marquez, N., Abbasalizad-Farhangi, M., Abbasi, M., Abd-Allah, F., et~al. (2019).
\newblock Mapping 123 million neonatal, infant and child deaths between 2000 and 2017.
\newblock {\em Nature}, 574(7778):353--358.

\bibitem[Chipman et~al., 1998]{chipman1998bayesian}
Chipman, H.~A., George, E.~I., and McCulloch, R.~E. (1998).
\newblock {Bayesian CART model search}.
\newblock {\em Journal of the American Statistical Association}, 93(443):935--948.

\bibitem[Chipman et~al., 2010]{chipman2010bart}
Chipman, H.~A., George, E.~I., and McCulloch, R.~E. (2010).
\newblock {BART: Bayesian additive regression trees}.
\newblock {\em The Annals of Applied Statistics}, 4(1):266--298.

\bibitem[Davies and Van Der~Laan, 2016]{davies2016optimal}
Davies, M.~M. and Van Der~Laan, M.~J. (2016).
\newblock {Optimal spatial prediction using ensemble machine learning}.
\newblock {\em The International Journal of Biostatistics}, 12(1):179--201.

\bibitem[Daw and Wikle, 2023]{daw2023reds}
Daw, R. and Wikle, C.~K. (2023).
\newblock {REDS: Random ensemble deep spatial prediction}.
\newblock {\em Environmetrics}, 34(1):e2780.

\bibitem[Dezeure et~al., 2015]{dezeure2015high}
Dezeure, R., B{\"u}hlmann, P., Meier, L., and Meinshausen, N. (2015).
\newblock High-dimensional inference: confidence intervals, p-values and r-software hdi.
\newblock {\em Statistical Science}, pages 533--558.

\bibitem[Diggle and Giorgi, 2019]{diggle2019model}
Diggle, P.~J. and Giorgi, E. (2019).
\newblock {\em Model-Based Geostatistics for Global Public Health: Methods and Applications}.
\newblock CRC Press.

\bibitem[Eddelbuettel, 2013]{eddelbuettel2013seamless}
Eddelbuettel, D. (2013).
\newblock {\em Seamless {R} and {C}++ Integration with {R}cpp}.
\newblock Springer.

\bibitem[Eddelbuettel and Balamuta, 2018]{eddelbuettel2018extending}
Eddelbuettel, D. and Balamuta, J.~J. (2018).
\newblock {Extending R with C++: a brief introduction to Rcpp}.
\newblock {\em The American Statistician}, 72(1):28--36.

\bibitem[Eddelbuettel and Fran{\c{c}}ois, 2011]{eddelbuettel2011rcpp}
Eddelbuettel, D. and Fran{\c{c}}ois, R. (2011).
\newblock {Rcpp: Seamless R and C++ Integration}.
\newblock {\em Journal of Statistical Software}, 40:1--18.

\bibitem[Fay and Herriot, 1979]{fay1979estimates}
Fay, R.~E. and Herriot, R.~A. (1979).
\newblock Estimates of income for small places: an application of {James-Stein} procedures to census data.
\newblock {\em Journal of the American Statistical Association}, 74(366a):269--277.

\bibitem[Friedman, 2001]{friedman2001greedy}
Friedman, J.~H. (2001).
\newblock {Greedy function approximation: a gradient boosting machine}.
\newblock {\em Annals of Statistics}, pages 1189--1232.

\bibitem[Fuglstad et~al., 2019]{Fuglstad19}
Fuglstad, G.-A., Simpson, D., Lindgren, F., and Rue, H. (2019).
\newblock Constructing priors that penalize the complexity of {G}aussian random fields.
\newblock {\em Journal of the American Statistical Association}, 114(525):445--452.

\bibitem[Georganos et~al., 2021]{georganos2021geographical}
Georganos, S., Grippa, T., Niang~Gadiaga, A., Linard, C., Lennert, M., Vanhuysse, S., Mboga, N., Wolff, E., and Kalogirou, S. (2021).
\newblock Geographical random forests: a spatial extension of the random forest algorithm to address spatial heterogeneity in remote sensing and population modelling.
\newblock {\em Geocarto International}, 36(2):121--136.

\bibitem[Gneiting and Raftery, 2007]{gneiting2007strictly}
Gneiting, T. and Raftery, A.~E. (2007).
\newblock Strictly proper scoring rules, prediction, and estimation.
\newblock {\em Journal of the American Statistical Association}, 102(477):359--378.

\bibitem[Goldstein et~al., 2015]{goldstein2015peeking}
Goldstein, A., Kapelner, A., Bleich, J., and Pitkin, E. (2015).
\newblock {Peeking inside the black box: Visualizing statistical learning with plots of individual conditional expectation}.
\newblock {\em Journal of Computational and Graphical Statistics}, 24(1):44--65.

\bibitem[G{\'o}mez-Rubio and Rue, 2018]{gomez2018markov}
G{\'o}mez-Rubio, V. and Rue, H. (2018).
\newblock Markov chain monte carlo with the integrated nested laplace approximation.
\newblock {\em Statistics and Computing}, 28(5):1033--1051.

\bibitem[Hubin and Storvik, 2016]{hubin2016estimating}
Hubin, A. and Storvik, G. (2016).
\newblock {Estimating the marginal likelihood with Integrated nested Laplace approximation (INLA)}.
\newblock {\em arXiv preprint arXiv:1611.01450}.

\bibitem[Kassie and Workie, 2019]{kassie2019exploring}
Kassie, G.~W. and Workie, D.~L. (2019).
\newblock Exploring the association of anthropometric indicators for under-five children in {E}thiopia.
\newblock {\em BMC Public Health}, 19(1):1--6.

\bibitem[Knorr-Held and Rue, 2002]{knorr2002block}
Knorr-Held, L. and Rue, H. (2002).
\newblock On block updating in {M}arkov random field models for disease mapping.
\newblock {\em Scandinavian Journal of Statistics}, 29(4):597--614.

\bibitem[Krueger et~al., 2020]{Krueger2020}
Krueger, R., Bansal, P., and Buddhavarapu, P. (2020).
\newblock {A new spatial count data model with Bayesian additive regression trees for accident hot spot identification}.
\newblock {\em Accident Analysis and Prevention}, 144:105623.

\bibitem[LeSage and {Kelley Pace}, 2007]{lesage07}
LeSage, J.~P. and {Kelley Pace}, R. (2007).
\newblock {A matrix exponential spatial specification}.
\newblock {\em Journal of Econometrics}, 140(1):190--214.
\newblock Analysis of spatially dependent data.

\bibitem[Lindgren et~al., 2011]{lindgren2011explicit}
Lindgren, F., Rue, H., and Lindstr{\"o}m, J. (2011).
\newblock {An explicit link between Gaussian fields and Gaussian Markov random fields: the stochastic partial differential equation approach}.
\newblock {\em Journal of the Royal Statistical Society: Series B (Statistical Methodology)}, 73(4):423--498.

\bibitem[Lindstr{\"o}m et~al., 2014]{lindstrom2014flexible}
Lindstr{\"o}m, J., Szpiro, A.~A., Sampson, P.~D., Oron, A.~P., Richards, M., Larson, T.~V., and Sheppard, L. (2014).
\newblock A flexible spatio-temporal model for air pollution with spatial and spatio-temporal covariates.
\newblock {\em Environmental and Ecological Statistics}, 21:411--433.

\bibitem[Macharia et~al., 2019]{macharia2019sub}
Macharia, P.~M., Giorgi, E., Thuranira, P.~N., Joseph, N.~K., Sartorius, B., Snow, R.~W., and Okiro, E.~A. (2019).
\newblock {Sub national variation and inequalities in under-five mortality in Kenya since 1965}.
\newblock {\em BMC Public Health}, 19(1):1--12.

\bibitem[Martins et~al., 2013]{martins2013bayesian}
Martins, T.~G., Simpson, D., Lindgren, F., and Rue, H. (2013).
\newblock Bayesian computing with inla: new features.
\newblock {\em Computational Statistics and Data Analysis}, 67:68--83.

\bibitem[Mei and Grummer-Strawn, 2007]{mei2007standard}
Mei, Z. and Grummer-Strawn, L.~M. (2007).
\newblock {Standard deviation of anthropometric Z-scores as a data quality assessment tool using the 2006 WHO growth standards: a cross country analysis}.
\newblock {\em Bulletin of the World Health Organization}, 85:441--448.

\bibitem[M{\"u}ller et~al., 2007]{muller2007spatially}
M{\"u}ller, P., Shih, Y.-C.~T., and Zhang, S. (2007).
\newblock {A spatially-adjusted Bayesian additive regression tree model to merge two datasets}.
\newblock {\em Bayesian Analysis}, 2(3):611--633.

\bibitem[Osgood-Zimmerman et~al., 2018]{osgood2018mapping}
Osgood-Zimmerman, A., Millear, A.~I., Stubbs, R.~W., Shields, C., Pickering, B.~V., Earl, L., Graetz, N., Kinyoki, D.~K., Ray, S.~E., Bhatt, S., et~al. (2018).
\newblock {Mapping child growth failure in Africa between 2000 and 2015}.
\newblock {\em Nature}, 555(7694):41--47.

\bibitem[Polson et~al., 2013]{polson2013bayesian}
Polson, N.~G., Scott, J.~G., and Windle, J. (2013).
\newblock Bayesian inference for logistic models using p{\'o}lya--gamma latent variables.
\newblock {\em Journal of the American statistical Association}, 108(504):1339--1349.

\bibitem[Rao and Molina, 2015]{rao2015small}
Rao, J.~N. and Molina, I. (2015).
\newblock {\em Small Area Estimation}.
\newblock John Wiley \& Sons.

\bibitem[Ren et~al., 2018]{ren2018}
Ren, Z., Zhu, J., Gao, Y., Yin, Q., Hu, M., Dai, L., Deng, C., Yi, L., Deng, K., Wang, Y., Li, X., and Wang, J. (2018).
\newblock {Maternal exposure to ambient PM10 during pregnancy increases the risk of congenital heart defects: Evidence from machine learning models}.
\newblock {\em Science of The Total Environment}, 630:1--10.

\bibitem[Rue et~al., 2009]{rue2009approximate}
Rue, H., Martino, S., and Chopin, N. (2009).
\newblock {Approximate Bayesian inference for latent Gaussian models by using integrated nested Laplace approximations}.
\newblock {\em Journal of the Royal Statistical Society: Series B (Statistical Methodology)}, 71(2):319--392.

\bibitem[Shi et~al., 2021]{shi2021digital}
Shi, T., Hu, X., Guo, L., Su, F., Tu, W., Hu, Z., Liu, H., Yang, C., Wang, J., Zhang, J., et~al. (2021).
\newblock {Digital mapping of zinc in urban topsoil using multisource geospatial data and random forest}.
\newblock {\em Science of the Total Environment}, 792:148455.

\bibitem[Spanbauer and Sparapani, 2021]{spanbauer2021nonparametric}
Spanbauer, C. and Sparapani, R. (2021).
\newblock {Nonparametric machine learning for precision medicine with longitudinal clinical trials and Bayesian additive regression trees with mixed models}.
\newblock {\em Statistics in Medicine}, 40(11):2665--2691.

\bibitem[Utazi et~al., 2018]{utazi2018high}
Utazi, C.~E., Thorley, J., Alegana, V.~A., Ferrari, M.~J., Takahashi, S., Metcalf, C. J.~E., Lessler, J., and Tatem, A.~J. (2018).
\newblock High resolution age-structured mapping of childhood vaccination coverage in low and middle income countries.
\newblock {\em Vaccine}, 36(12):1583--1591.

\bibitem[Uwiringiyimana et~al., 2022]{uwiringiyimana2022bayesian}
Uwiringiyimana, V., Osei, F., Amer, S., and Veldkamp, A. (2022).
\newblock {Bayesian geostatistical modelling of stunting in Rwanda: risk factors and spatially explicit residual stunting burden}.
\newblock {\em BMC Public Health}, 22(1):1--14.

\bibitem[Zeraatpisheh et~al., 2022]{ZERAATPISHEH2022105723}
Zeraatpisheh, M., Garosi, Y., {Reza Owliaie}, H., Ayoubi, S., Taghizadeh-Mehrjardi, R., Scholten, T., and Xu, M. (2022).
\newblock Improving the spatial prediction of soil organic carbon using environmental covariates selection: A comparison of a group of environmental covariates.
\newblock {\em CATENA}, 208:105723.

\end{thebibliography}

\end{document}